\documentclass[prl,aps,twocolumn,eclepsf,amsmath,amssymb,showpacs]{revtex4}  % DON'T CHANGE(two colme size)

\usepackage[dvips]{graphicx}% Include figure files
\usepackage{dcolumn}% Align table columns on decimal point
\usepackage{bm}% bold math

\begin{document}                % INITIALIZE - DONT CHANGE
\draft
\title{Effects of next-nearest-neighbor hopping $t^{\prime}$ on the electronic structure of cuprates}

\author{K.~Tanaka, T.~Yoshida, A.~Fujimori, D.H.~Lu$^{\dagger}$, Z.-X.~Shen$^{\dagger}$, X.-J.~Zhou$^{\dagger}$$^{\ast}$, H.~Eisaki$^{\dagger}$, Z.~Hussain$^{\ast}$, S.~Uchida, Y.~Aiura$^{\flat}$, K.~Ono$^{\sharp}$, T.~Sugaya$^{\ddagger}$, T.~Mizuno$^{\ddagger}$,
and I.~Terasaki$^{\ddagger}$}

\address{Department of Physics and Department of Complexity Science
  and Engineering, University of Tokyo, Tokyo, 113-0033, Japan\\
$^{\dagger}$Department of  Applied Physics and Stanford
Synchrotron Radiation Laboratory, Stanford University, Stanford,
CA 94305, USA\\ $^{\ast}$ Advanced Light Source, Lawrence Berkely
National Lab, Berkeley, CA 94720, USA\\ $^{\flat}$ National
Institute for Advanced Industrial Science and Technology (AIST),
Tsukuba,
305-8568, Japan\\ $^{\sharp}$ Photon Factory, IMSS, High Energy Accelerator Research Organization, Tsukuba, 305-0801, Japan\\
$^{\ddagger}$Department of Applied Physics, Waseda University,
Tokyo 169-8555, Japan}

%\date{Received {\today}}
\date{\today}

\begin{abstract}                % DON'T CHANGE THIS LINE
Photoemission spectra of underdoped and lightly-doped
Bi$_{2-z}$Pb$_z$Sr$_2$Ca$_{1-x}${\it R}$_{x}$Cu$_2$O$_{8+y}$ ($R=$
Pr, Er) (BSCCO) have been measured and compared with those of
La$_{2-x}$Sr$_x$CuO$_4$ (LSCO). The lower-Hubbard band of the
insulating BSCCO, like Ca$_2$CuO$_2$Cl$_2$, shows a stronger
dispersion than La$_2$CuO$_4$ from ${\bf k}\sim$($\pi/2,\pi/2$) to
$\sim$($\pi,0$). The flat band at ${\bf k}\sim$($\pi,0$) is found
generally deeper in BSCCO. These observations together with the
Fermi-surface shapes and the chemical potential shifts indicate
that the next-nearest-neighbor hopping $|t^{\prime}|$ of the
single-band model is larger in BSCCO than in LSCO and that
$|t^{\prime}|$ rather than the super-exchange $J$ influences the
pseudogap energy scale.

\end{abstract}
 \maketitle

%\section{Introduction}
Since the discovery of the high-temperature superconductivity in
La$_{2-x}$Ba$_x$CuO$_4$, many families of high-$T_c$ cuprates have
been synthesized. Common features are that they have the
two-dimensional CuO$_2$ planes and a similar phase diagram as a
function of hole doping. This has naturally lead most of studies
to emphasize the common features of the cuprate electronic
structures rather than emphasizing differences among them. On the
other hand, there are differences among the different families of
cuprates such as the significant variation in the magnitude of the
superconducting gap and the critical temperature ($T_c$) at
optimal doping, $T_{c, {\rm max}}$. A systematic investigation of
the differences between the different families of cuprates may
enable us to understand the origin of the different $T_{c, {\rm
max}}$'s and eventually the mechanism of superconductivity. So
far, some studies have focused on the material dependence from
empirical points of view. In an early work, Ohta \textit{et
al.}~\cite{Madelung} proposed the differences in the position of
the apical oxygen and the resulting differences in the Madelung
potentials as the origin of the different $T_{c, {\rm max}}$'s.
Feiner \textit{et al.}~\cite{Feiner} proposed that the $p_z$
orbital of the apical oxygen hybridizing with the $d_{3z^2-r^2}$
orbital of Cu and the $p_{x,y}$ orbitals of the in-plane oxygen
affects the next-nearest-neighbor hopping parameter $t^{\prime}$
in the single-band model description of the CuO$_2$ plane, and
thereby $T_{c, {\rm max}}$ in the context of the van Hove
singularity scenario~\cite{Dagotto}. Those differences between the
cuprate families may affect the stability of the Zhang-Rice
singlet~\cite{Madelung}, instability toward charge
stripes~\cite{Fleck} and so on, and hence $T_{c, {\rm max}}$.
Recently, Pavarini \textit{et al.}~\cite{Andersen} have
demonstrated the correlation between $t^{\prime}$ (of the bonding
band for multilayer cuprates) and $T_{c, {\rm max}}$ from their
tight-binding model analysis of the first-principles band
structures of numerous high-$T_c$ cuprates. For the differences in
$T_{c, {\rm max}}$, the various degrees of disorder has also been
considered important~\cite{Eisaki}.

In the present work, on the basis of photoemission data, we focus
on differences in the electronic structure of the cuprates such as
the band dispersion of the parent insulator and the doped
compounds as well as the Fermi surface shape between
La$_{2-x}$Sr$_x$CuO$_4$ (LSCO) and
Bi$_2$Sr$_2$CaCu$_2$O$_{8+\delta}$ (BSCCO). We have found that
lightly-doped and underdoped BSCCO show a stronger band dispersion
along the ``underlying Fermi surface" than its counter part in
LSCO. Given that $J$ does not change much between the two families
($J_{\textrm{LSCO}}$$\sim139$ meV and
$J_{\textrm{BSCCO}}$$\sim127$ meV from two-magnon Raman
scattering~\cite{Raman_BSCCO,Raman_LSCO,Raman_YBCO} and magnetic
neutron scattering~\cite{neutron_YBCO}), we attribute the observed
differences to the variation in $t^{\prime}$, a finding consistent
with the band structure estimates of $t^{\prime}$~\cite{Andersen}
and the $t$-$J$ model calculation on the impact of $t^{\prime}$ on
the electronic structure around ${\bf
k}\sim$($\pi,0$)~\cite{SCOC3}.

So far, photoemission studies of LSCO have covered a wide
composition range from the lightly-doped to overdoped regions and
systematic data are available for the evolution of the
pseudogap~\cite{Ino_pseudogap}, Fermi
surface~\cite{Ino_Fermisurf1,Ino_Fermisurf2,XJZhou}, band
dispersion~\cite{Ino_Fermisurf2,Yoshida_lightly} and chemical
potential shift~\cite{Ino_chemipote}. Although BSCCO has been
extensively studied by angle-resolved photoemission spectroscopy
(ARPES) owing to its stable cleavage surfaces in an ultra high
vacuum, the available range of hole concentration has been largely
limited to $\delta=0.10$-$0.17$. Recently, high quality single
crystals of heavily underdoped BSCCO were synthesized by
rare-earth ($R$) substitution for Ca and the doping dependence of
thermodynamic and transport properties have been systematically
studied~\cite{Terasaki,sample,Pbsample}. The present study was
made possible by the availability of such deeply underdoped BSCCO
samples.

Single crystals of Bi$_{1.2}$Pb$_{0.8}$Sr$_2$ErCu$_2$O$_8$ and
Bi$_2$Sr$_2$Ca$_{1-x}${\it R}$_{x}$Cu$_2$O$_{8+y}$ ($R=$ Pr, Er)
were grown by the self-flux method. X-ray diffraction showed no
trace of impurity phases. Details of the sample preparation are
given elsewhere~\cite{sample,Pbsample}. The hole concentration
${\delta}$ per Cu atom was determined using the empirical
relationship between ${\delta}$ and the room-temperature
thermopower~\cite{thermopower}. The $\delta$ and $T_{c}$ of the
measured samples are listed in Table.\ 1. The $x=0.5$ and 1.0 Er
samples are antiferromagnetic (AF) insulators. The Laue patterns
of the Pb doped samples showed no superlattice modulation of the
Bi-O layers, which eliminated superstructure signals in ARPES
spectra. Single crystals of LSCO were grown by the
traveling-solvent floating-zone method. The $T_{c}$ of $x=0.07$,
0.10, 0.15 and 0.22 samples were 14, 29, 41 and 20 K,
respectively, and $x=0.00$, 0.03 samples were
non-superconducting~\cite{Yoshida_lightly}.

\begin{table}[tb]
\caption{Chemical compositions, hole concentration $\delta$ and
$T_c$ of BSCCO samples studied in the present work.}
\label{table1}
\begin{center}
\begin{ruledtabular}
\begin{tabular}{ccc}
Bi$_2$Sr$_2$Ca$_{1-x}R_{x}$Cu$_2$O$_8$&$\delta$&$T_c$(K)\\
\hline
$R$ = Er, $x=1$ & 0.025 & -\\
$R$ = Er, $x=0.5$ & 0.05 & -\\
$R$ = Er, $x=0.1$ & 0.135 & 87\\
$R$ = Pr, $x=0.43$ & 0.1 & 48\\
$R$ = Pr, $x=0.25$ & 0.135 & 88\\
$R$ = Pr, $x=0.1$ & 0.17 & 86\\
\hline
Bi$_{1.2}$Pb$_{0.8}$Sr$_2$ErCu$_2$O$_8$ & 0.04 & -\\
\end{tabular}
\end{ruledtabular}
\end{center}
\end{table}

ARPES measurements of BSCCO were carried out at beamline 5-4 of
Stanford Synchrotron Radiation Laboratory (SSRL). Incident photons
had an energy of ${\it h}{\nu}=19$ eV. A SCIENTA SES-200 analyzer
was used in the angle mode with the total energy and momentum
resolutions of $\sim14$ meV and $\sim0.25^\circ$, respectively.
Samples were cleaved \textit{in situ} under an ultrahigh vacuum of
$10^{-11}$ Torr, and were cooled down to $\sim10$ K. The position
of the Fermi level ($E_{\rm{F}}$) was calibrated with gold
spectra. ARPES measurements with ${\it h}{\nu}=30-60$ eV at $85$ K
were performed at beamline BL-1C of Photon Factory (PF) using an
ARUPS-10 analyzer. The overall energy resolution varied from $130$
to $150$ meV. ARPES measurements of LSCO were carried out at
BL10.0.1.1 of Advanced Light Source (ALS), using incident photons
of $55.5$ eV at $20$ K as described
elsewhere~\cite{Yoshida_lightly}. Angle-integrated photoemission
spectroscopy (AIPES) measurements of BSCCO samples were carried
out using the He {\footnotesize I} resonance line (${\it
h}{\nu}=21.2$~eV) with an OMICRON 125EA analyzer. The samples were
cleaved \textit{in situ} and measured at $\sim7$ K with the energy
resolution of $\sim25$ meV.

\begin{figure}[tb]
\begin{center}
\includegraphics[width=90mm]{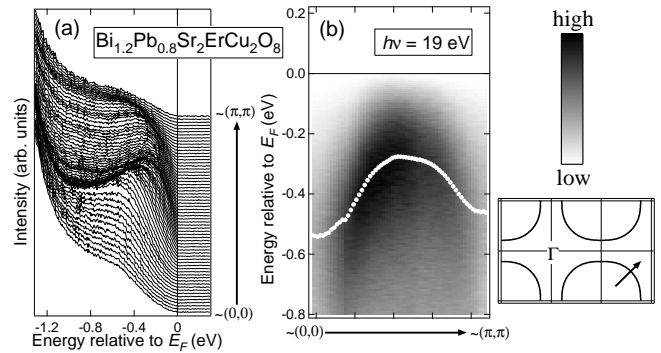}
\caption{ARPES spectra of insulating
Bi$_{1.2}$Pb$_{0.8}$Sr$_2$ErCu$_2$O$_8$ along the
(0,0)-($\pi,\pi$) direction. (a) EDC's, (b) Intensity plot in the
$E$-$k$ plane. The white circles indicate the energy of the
maximum curvature in the EDC's.}
\end{center}
\end{figure}

Figure 1 shows the ARPES spectra of insulating
Bi$_{1.2}$Pb$_{0.8}$Sr$_2$ErCu$_2$O$_8$ ($\delta\sim0.04$) along
the diagonal $(0,0)$-$(\pi,\pi)$ direction in the second Brillouin
zone (BZ). The figure shows a single dispersive feature
corresponding to the lower Hubbard band, which moves closest to
$E_{\rm{F}}$ at $\sim$($\pi/2,\pi/2$). There is no sharp peak
crossing $E_{\rm{F}}$. This is contrasted with LSCO of similar
doping level where a tiny but sharp peak crosses
$E_{\rm{F}}$~\cite{Yoshida_lightly}, and is consistent with the
insulating behavior of the present compound~\cite{Pbsample}. (Note
that LSCO with $x\sim0.03$ shows metallic behaviour at $T>100$ K.)

Figure 2 shows ARPES spectra along the ``underlying Fermi surface"
~\cite{Ronning}. One can again see a single dispersive feature
between $-0.6$ and $-0.2$ eV, as in the case of
Ca$_2$CuO$_2$Cl$_2$ (CCOC) and
Sr$_2$CuO$_2$Cl$_2$~\cite{Ronning,SCOC1,SCOC3}. In Fig.\ 2(c), we
have plotted the peak position of the spectra marked in Fig.\ 2(a)
referenced to the binding energy of the peak at ($\pi/2,\pi/2$)
against $|{\cos{k_xa}-\cos{k_ya}}|/2$. The nearly straight line
shows approximately $d_{x^2-y^2}$-like gap anisotropy on the
underlying Fermi surface~\cite{d_wave}. Combining Figs.\ 1(b) and
2(b), one can conclude that the band dispersion in the insulating
BSCCO is nearly isotropic around ($\pi/2,\pi/2$). Figure 2(c)
shows that the total dispersional width in the insulating BSCCO is
comparable to that in CCOC but is larger than that in undoped LSCO
(La$_2$CuO$_4$) by a factor of $\sim1.7$. Here, it should be
cautioned that spectra near ($\pi,0$) of BSCCO may be influenced
by possible bilayer splitting and that the intensity of the
bonding band (BB) and antibonding band (AB) show different ${\it
h}{\nu}$-dependence~\cite{Fink}. In order to check this
possibility, we measured the photon energy dependence of the
spectra of heavily underdoped samples ($\delta\sim0.025$ and
$0.06$) at ($\pi,0$) from ${\it h}{\nu}=30$~eV to $60$~eV with
$5$~eV photon energy interval, where the relative intensities of
the BB and AB are expected different. We did not find appreciable
photon-energy dependence in the line shape and the peak position
from ${\it h}{\nu}=19$~eV. We therefore consider the impact of
bilayer splitting on the ($\pi,0$) electronic structure is small,
partly because the ($\pi,0$) state is already pushed considerably
below the Fermi level.

\begin{figure}[tb]
\begin{center}
\includegraphics[width=90mm]{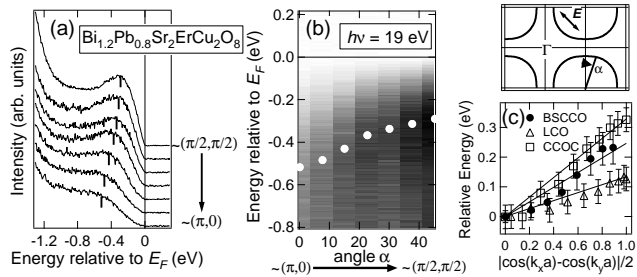}
\caption{ARPES spectra of Bi$_{1.2}$Pb$_{0.8}$Sr$_2$ErCu$_2$O$_8$
along the ``underlying Fermi surface" in the second BZ. (a) EDC's.
(b) Intensity plot in the $E$-$k$ plane. Vertical bars in (a) and
the white circles in (b) indicate the points of maximum curvature
in the EDC's. (c) Peak positions referenced to the binding energy
of the peak at ($\pi/2,\pi/2$) plotted against
$|{\cos{k_xa}-\cos{k_ya}}|/2$. Also plotted are data for
La$_2$CuO$_4$ (LCO) and Ca$_2$CuO$_2$Cl$_2$
(CCOC)~\protect\cite{Ronning}.}
\end{center}
\end{figure}

In order to interpret the band dispersion in the parent insulator
within the single-band description, we first consider the Hubbard
($U$-$t$) model or the $t$-$J$ model, where $J=4t^2/U$, and $t$,
$J$, and $U$ are the nearest neighbor hopping matrix element, the
AF super-exchange coupling constant, and the on-site Coulomb
energy, respectively. These models can explain the experimental
band dispersion from (0,0) to ($\pi,\pi$) because its width is
predicted to be $\sim2.2J$ ($\sim0.28$ eV)~\cite{SCOC1}, however,
they predict nearly the same peak energies for ($\pi/2,\pi/2$) and
($\pi,0$), disagreeing with the strong dispersion along the
underlying Fermi surface in the insulating BSCCO. According to an
extended version of the Hubbard model or the $t$-$J$ model, i.e.,
the $t$-$t^{\prime}$-$t^{\prime\prime}$-$U$ model or the
$t$-$t^{\prime}$-$t^{\prime\prime}$-$J$ model, which takes into
account the hopping to the second and third nearest neighbors
through $t^{\prime}$ and $t^{\prime\prime}$, the strong dispersion
from ($\pi/2,\pi/2$) to ($\pi,0$) can be realized by a sizeable
$t^{\prime}$~\cite{SCOC3}. This implies a significantly larger
value of $|t^{\prime}|$ in BSCCO than in LSCO.

\begin{figure}[tb]
\begin{center}
\includegraphics[width=90mm]{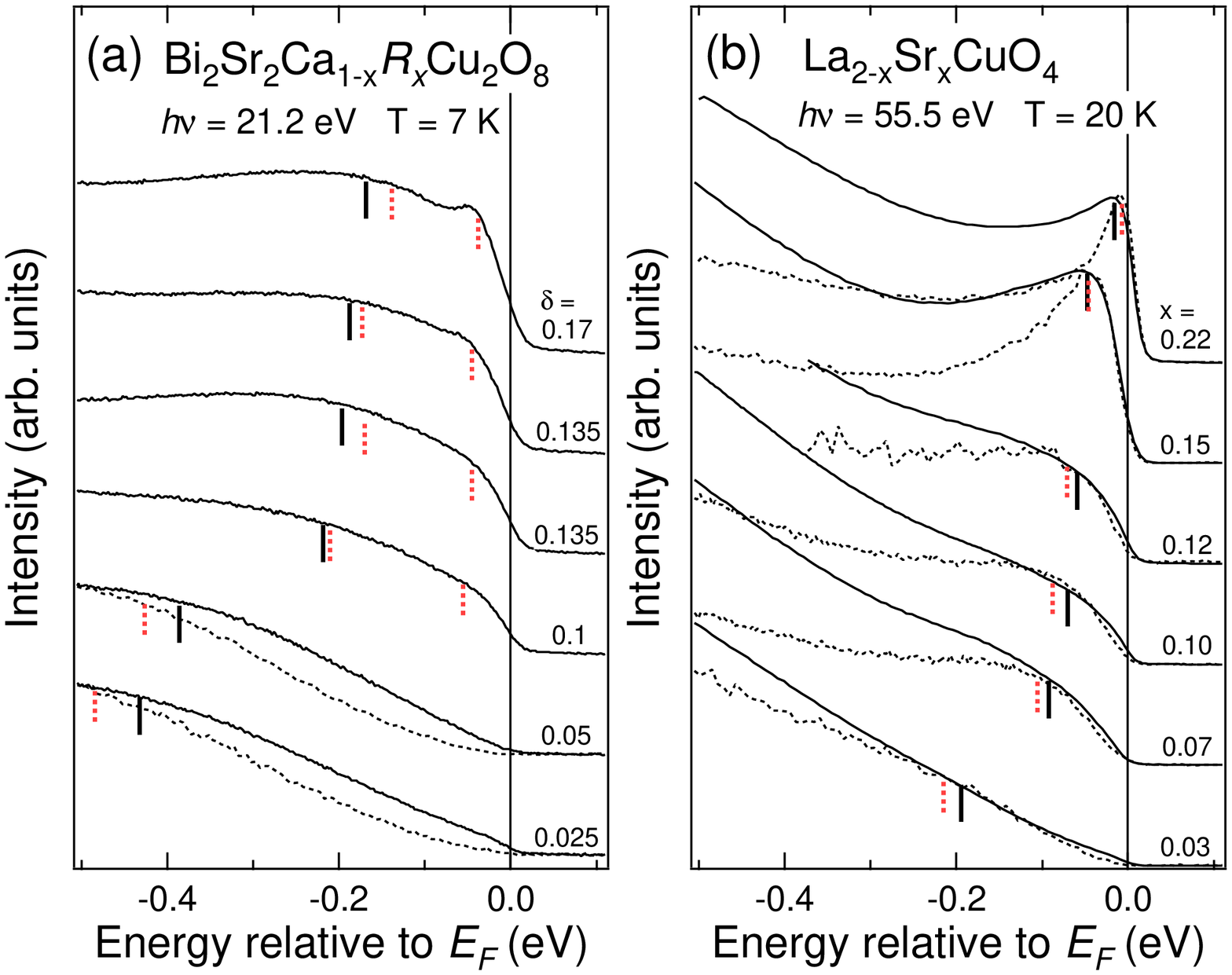}
\caption{AIPES spectra (solid curve) and ARPES spectra at
($\pi,0$) (dashed curve) for various doping levels in BSCCO (a)
and LSCO (b). Solid vertical bars indicate the point of maximum
curvature and the dashed vertical bars mark the position of the
flat band $E(\pi,0)$ (BB and AB at higher and lower binding
energies, respectively, for BSCCO). ARPES data for BSCCO with
${\delta}{\geq}0.1$ was taken
from~\cite{peakdiphump,superfluid,hump}}
\end{center}
\end{figure}

In order to see further differences between BSCCO and LSCO, we
show in Fig.\ 3 the ARPES spectra at ($\pi,0$) (dashed curves) and
AIPES spectra (solid curves) of BSCCO and LSCO. The AIPES spectra
of LSCO have been obtained by integrating ARPES data within the
second BZ. In the overdoped (${\delta}=0.17$) BSCCO sample, one
can see a well-known peak-dip-hump structure as observed in the
ARPES spectra near the ($\pi,0$)
point~\cite{peakdiphump,superfluid}. The intensity of the peak at
$\sim -40$ meV decreases with decreasing hole concentration or
with increasing temperature (not shown). The dashed vertical bars
in Fig.\ 3 mark the position of the flat band $E(\pi,0)$ (for
BSCCO, BB and AB at higher and lower binding energies,
respectively) in the ARPES spectra at $(\pi,0)$. The solid
vertical bars mark the point of the maximum curvature in the
second derivatives of the AIPES spectra~\cite{Ino_pseudogap}, also
representing $E(\pi,0)$ in LSCO and the energy position of BB in
BSCCO. In Fig.\ 4, those $|E(\pi,0)|$ values for LSCO and BSCCO
and the average energy positions of BB and AB for BSCCO,
$\overline{|E(\pi,0)|}$, are plotted. One can see that
$\overline{|E(\pi,0)|}$ in BSCCO is larger by a factor of $\sim2$
than $|E(\pi,0)|$ in LSCO. Note that the magnitude of the small
pseudogap, i.e., the binding energy of the leading edge position
is also larger in BSCCO than in LSCO by a factor of
$\sim2$~\cite{Ino_Fermisurf2}. On the other hand, the $J$ values
are almost common between different families of
cuprates~\cite{Raman_BSCCO,Raman_LSCO,Raman_YBCO,neutron_YBCO}.
The present observation therefore suggests that the
``band-structure" effect represented by $t^{\prime}$ has important
effect on the magnitude of the large pseudogap. It should be noted
that the energy position of the flat band was shown to become
deeper with increasing $|t^{\prime}|$ according to the
$t$-$t^{\prime}$-$t^{\prime\prime}$-$J$ model
calculations~\cite{SCOC3,Prelovsek}. As for the shape of the Fermi
surface, since the flat band at $(\pi,0)$ especially of BB is
deeper in BSCCO, the crossing point along the $(0,0)$-$(\pi,\pi)$
line becomes closer to (0,0) and that along the
$(\pi,0)$-$(\pi,\pi)$ line becomes closer to $(\pi,\pi)$, leading
to the more ``square-like" hole Fermi surface centered at
$(\pi,\pi)$ compared to the ``diamond-like" hole Fermi surface in
optimally-doped LSCO~\cite{Fermisurface,Prelovsek}. Recently, the
chemical potential shift as a function of doping was found to be
faster in BSCCO than in LSCO~\cite{harima}, which can also be
explained by a larger value of $|t^{\prime}|$ in BSCCO based on
exact diagonalization studies of the $t$-$t^{\prime}$-$J$
model~\cite{chemipote}.

\begin{figure}[tb]
\begin{center}
\includegraphics[width=70mm]{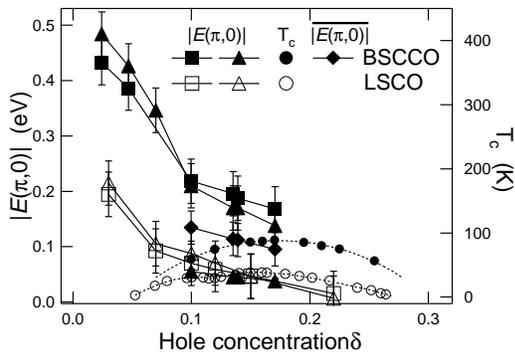}
\caption{Doping dependence of the flat band position $E(\pi,0)$ in
BSCCO and LSCO determined by the second derivatives of AIPES
spectra (squares) and ARPES spectra (triangles). The average
energy position of BB and AB, $\overline{|E(\pi,0)|}$, for BSCCO
(diamond) and $T_{c}$ (circle) are also shown.}
\end{center}
\end{figure}

Figure 4 suggests that $|E(\pi,0)|$ of LSCO and
$\overline{|E(\pi,0)|}$ of BSCCO are scaled by $T_{c, {\rm max}}$.
This implies possible relationship between $T_{c, {\rm max}}$ and
$t^{\prime}$, as has been suggested in several different
contexts~\cite{Feiner,Andersen}. $T_{c, {\rm max}}$ in the
high-$T_c$ superconductors is determined by the block layer (and
hence the position of the apical oxygen) as well as the number of
CuO$_2$ planes. It has been found from first-principles
calculations that the values of $t^{\prime}$ and
$t^{\prime\prime}$ are different between different families of
cuprates~\cite{Andersen}, while the other parameters are rather
material-independent; $U\sim3$ eV, $t\sim0.3$ eV and
$J=4t^2/U\sim0.1$ eV. Therefore, it is quite natural to consider
that the difference in $t^{\prime}$ (and probably
$t^{\prime\prime}$) rather than $J$ strongly affects the $T_{c,
{\rm max}}$. Ohta \textit{et al.}~\cite{Madelung} argued that the
Cu atom-apical oxygen distance affects the Madelung potential
difference between the apical oxygen and the oxygen in the plane
and thereby influences the stability of the Zhang-Rice singlet
relative to the \textit{B}$_{1\textrm{g}}$ triplet where the hole
enters the apical oxygen $p_z$ orbital. According to Fleck
\textit{et al.}~\cite{Fleck} and Tohyama \textit{et
al.}~\cite{ladder}, the stability of charge stripes increases with
decreasing $|t^{\prime}/t|$ but does not depend on $J/t$,
consistent with the observation that LSCO is closer to the
instability toward stripe-type spin-charge
ordering~\cite{stripe1}. It has also been proposed that the Fermi
surface shape itself, which is influenced by $t^{\prime}$ and
$t^{\prime\prime}$, is important to increase $T_{c, {\rm
max}}$~\cite{Andersen}. More theoretical studies are needed to
identify a microscopic mechanism in which larger $|t^{\prime}|$
leads to higher $T_{c, {\rm max}}$.

In conclusion, we have identified several differences between the
electronic structures of BSCCO and LSCO, all of which can be
explained by the larger value of $|t^{\prime}|$ in BSCCO than in
LSCO. In order to see whether there is indeed correlation between
$T_{c, {\rm max}}$ and $|t^{\prime}|$, further systematic studies
on other materials (such as YBCO, Tl-bases cuprates,
\textrm{etc.}) are highly desirable.

We acknowledge technical help by N.\ P.\ Armitage and K.\ M.\
Shen, and enlightening discussion with T.\ Tohyama, D.\ L.\ Feng,
I.\ Dasgupta, E.\ Parvarini and O.\ K.\ Andersen. This work was
supported by a Grant-in-Aid for Scientific Research in Priority
Area ``Novel Quantum Phenomena in Transition-Metal Oxides" from
the Ministry of Education, Culture, Sports, Science and Technology
of Japan, the New Energy and Industrial Technology Development
Organization (NEDO) and a US-Japan Joint Research Project from the
Japan Society for the Promotion of Science. SSRL is operated by
the DOE Office of Basic Energy Science Divisions of Chemical
Sciences and Material Sciences. ALS is operated by the DOE Office
of Basic Energy Science, Division of Material Science. This work
was also performed under the approval of the Photon Factory
Program Advisory Committee (Proposal No.\ 2002G174).

\newpage

\end{document}